\documentstyle[fleqn]{tp}

\input psfig

\def\mpc{\mbox{Mpc}}
\def\kpc{\mbox{kpc}}
\def\deg{\mbox{deg}}
\def\yr{\mbox{yr}}
 \def\simlt{\lower.5ex\hbox{$\; \buildrel < \over \sim \;$}}
  \def\simgt{\lower.5ex\hbox{$\; \buildrel > \over \sim \;$}}

\begin{document}

\twocolumn[
\title{The Detection of The Gravitational Lensing of Supernovae}
\author{R. Benton Metcalf\\
{\it Physics and Astronomy Departments, University of California,} \\ {\it Berkeley, CA 94720 USA}}
\vspace*{16pt}   

ABSTRACT.\
The weak gravitational lensing of high redshift type Ia supernovae
has the potential of probing the structure of matter on galaxy 
halo scales.  This is complementary to the weak lensing of galaxies 
which probes structure of larger scales.
There are already several organized searches for these supernovae being 
carried out for the purposes of cosmological parameter 
estimation.  A method is proposed for extracting from future supernovae 
data information on lensing and the structures responsible.  This method 
utilizes the correlations between SN luminosities and foreground galaxies.
Simulations of the lensing and uncertainties will be presented.
It is found that with a hundred supernovae or more at $z\simgt 1$ or larger, 
significant measurements of the mass, shape and extent of dark matter halos 
could be made if they contain a significant proportion of the matter in the universe.

\endabstract]

\markboth{RB Metcalf}{Gravitational Lensing of SNe}

\small

\section{Introduction}

Two of the most active and promising fields of observational cosmology today are the 
search for weak gravitational lensing by large scale structure (LSS) 
and the search for high redshift supernovae (SNe) for the purposes of determining 
the luminosity distance-redshift relation.  These two subjects are discussed 
elsewhere in this proceedings (Hook 1998, Schneider 1998 and Bartelmann 1998, 
these proceedings).  Here I will discuss another use for high redshift 
SNe, and one that is complementary to weak lensing measures of LSS.

Others have addressed the lensing of SNe in the past.  These investigations
have concentrated on either microlensing (Linder, Schneider \& Wagoner 1988) or the effect 
lensing has on uncertainties in cosmological parameter estimation 
(Frieman 1997, Wambsganss {\it et al.} 1997, Kantowski 1998, Holz 1998).  Here I address the problem of 
detecting lensing and recovering information about the lensing structures from future SN data.

A little review of the SN searches will be necessary.  The most important 
point here is that after corrections type Ia SNe have been shown empirically to 
be good standard candles with peak luminosities of $M_B \sim -19$.  A variance 
of $\Delta M_{V,B} \simeq 0.12\mbox{~mag}$ in the corrected peak magnitudes has been 
achieved using $V$ and $B$ bands ( Reiss, Press \& Kirshner 1996 ) and a 
$\Delta M_{B} \simeq 0.17\mbox{~mag}$ using one color (Hamuy {\it et al.} 1996).
It should be noted that these numbers are only approximations of the true 
intrinsic variation in peak luminosities because they are based on the statistics 
of just a few SNe (20 and 18, respectively).  The technique for correcting the peak luminosities may 
improve in the next few years as more SNe are found and additional information such as 
spectral features are incorporated.  At present over 100 type Ia SN have been 
observed, most of them 
below $z\simeq 0.5$.  Searches can now reliably discover about 10 SNe in a 
night of observing ($\sim 3 \mbox{~deg}^2$).  For more details on this subject see 
Hook in these proceedings and references there in.

In the next section some motivations for considering the lensing SNe will 
be given. In section~3 simulations of the lensing are described and in section~4 
these simulations are used in Monte Carlo calculations to estimate the accuracy 
of future experiments.

\section{Why Supernovae?}
It is reasonable to ask the question: Why try to detect the lensing of SNe 
when you can use galaxies which are much more numerous and can be observed at 
higher redshifts?  The answer is that the lensing of SNe and galaxies probe 
different size scales in the lensing structure.

Unfortunately we do not know the intrinsic properties such as ellipticities and 
luminosities of galaxies at $z \simgt 1$.  Not only do we not know the properties 
of individual galaxies, but because of galaxy evolution and the fact that galaxies 
are lensed we also do not know the intrinsic statistical distribution of any 
observable properties that are affected by lensing.  For example we do not know 
the rms ellipticity of unlensed galaxies at $z = 2$.  As a result galaxies taken 
individually can give us no information about lensing.  To search for lensing we 
must look for correlations amongst the images of galaxies - background and 
foreground or background and background - which are assumed to not exist in the 
real galaxies.\footnote{An exception to this is the case of strong lensing 
were the distortion in the image is very large.}
The necessity of using correlations between galaxies makes studies of the lensing 
of galaxies insensitive to scales below roughly the mean separation between galaxies 
even though these scales probably dominate the lensing of individual galaxies.  Noise 
increases this minimum scale to $\simgt 1'$.

Supernovae type Ia on the other hand are presumably the same objects today as they 
where at any redshift excepting the possibility of evolution in progenitor 
properties.  
By observing low redshift SNe we can determine the unlensed statistical properties 
of these sources - i.e. their peak luminosity distribution - and then take each 
high redshift SNe to be an independent measure of the magnification at a point which 
in general will be a larger signal than the magnification averaged over some area.

There is another reason why the lensing of SNe should give a higher 
signal.  Matter on small scales is expected to be 
centrally concentrated in clumps.  The shear, $\gamma(r)$, is a ``polar'' (a vector 
like object that rotates according to $2\theta$ instead of $\theta$) so it has 
a direction.  The influence of different mass clumps on a particular galaxy image 
can partially cancel.  The shear will then add like a random walk.  This is important 
for galaxy-galaxy lensing which attempts to measure the shape of galaxy halos.  This 
technique has been used to measure the mass within $\sim 30 h^{-1} \kpc$, but does 
not put strong constraints on the shape of halos (see Hudson {\it et al.} 1998 
and references there in).  In the weak lensing limit the magnification 
$\mu(r)-1 \propto \Sigma(r)$, where $\Sigma(r)$ is the projected surface density of 
the clump.  Magnification has no cancellations.  

Because the variance in the pointwise magnification is larger then the variance in 
the smoothed magnification it might be hoped that SNe will make up for their small 
numbers.  This is the question I wish to address in the rest of this paper.

\section{Simulations of Lensing}
To simulate the results of possible future observations a model for the 
lensing of point sources must first be made.  In the weak lensing limit 
the magnification can be written in terms of the convergence, $\kappa(\theta)$, 
$\mu(\theta) \cong 1+2\kappa(\theta)$.  Then the change in a sources' magnitude is 
\begin{equation}
\Delta m=2.5 \log(\mu(\theta)) \cong 2.17\kappa(\theta).
\end{equation}

\begin{figure*}
\vspace*{-1cm}
\centering\mbox{\psfig{figure=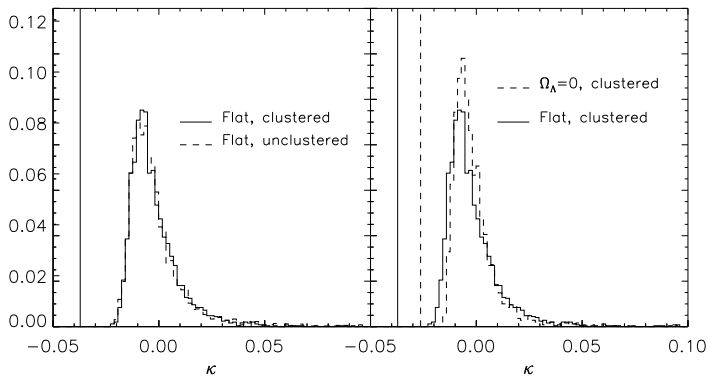,height=7cm}}
\vspace*{-1cm}
\caption[]{This is the distribution of convergences, $\kappa$, for 
SNe at $z=1$ in $\Omega_o=0.3$ universes, both open and $\Omega_o+\Omega_\Lambda=1$.  
The vertical lines on the left sides of the plots demark the ``empty beam'' 
solutions as explained in the text.}
\label{fig:1}
\end{figure*}

In this paper I will only be interested in the mass correlated with observable 
galaxies (for other possibilities see Metcalf 1998).  It will be assumed that 
this mass is in discrete halos surrounding each galaxy.  \footnote{The term 
``halos'' is perhaps a little over restrictive because what is really measured 
is the correlation between galaxies and mass.}  The convergence due to one halo is
\begin{equation}
\kappa_h(r_\perp)=\frac{\Sigma_h(r_\perp)-\Sigma_b}{\Sigma_c(z_h,z_{SN})} 
\equiv \tilde{\kappa}_h-\kappa_b
\end{equation}
The second term, $\Sigma_b$, accounts for the compensating effect of the background 
density.  This term is often left out of this expression when considering the lensing 
by clusters because it is small and shear maps are insensitive to a uniform offset.  
In the present case it will be important.  The convergence calculated without 
this term is $\tilde{\kappa}_h$ and $\Sigma_c$ is the critical density.  

The value of $\kappa_b$ depends on what distance is used to propagate the light 
between halos.  Here the usual Robertson-Walker distance will be used and $\kappa_b$ 
will be set so that $\langle \kappa \rangle=0$ and so the mean luminosity-redshift 
distance will agree with Robertson-Walker.  This can be done experimentally.  If the 
lensing is sufficiently weak the total magnification of a SN
will be the sum of the magnifications of each halo.
\begin{equation}
\kappa_{SN}=\sum_{halos} \left( \tilde{\kappa}_h - \langle \tilde{\kappa}_h \rangle_{SN} \right)
\label{kappa}
\end{equation}

The dark halo of each galaxy is modeled with the density profile
\begin{equation}
\rho(r)=\frac{V_c^2}{4\pi G r^2}\left[ \left(\frac{r}{r_s}\right)^\gamma +1 \right]^{-1}.
\label{halo}
\end{equation}
This is just an isothermal sphere sewn to a power law at a scale length $r_s$.
The central velocity is taken to be related to the luminosity of the galaxy through 
the Tully-Fisher relation, $V_c = V_f(L/L_f)^\beta$.  A Schechter function fitted 
to local galaxies is used for the luminosity function.  This model assumes that 
the halos were formed sometime before the SN went off and that there has not 
been substantial merging since that time.
For $\gamma \simlt 2$ a cutoff radius is usually needed so that the halos do not over 
fill the universe.  This additional parameter can be expressed as the total mass in 
halos relative to the critical mass, $\Omega_h$.

The clustering of halos is done by running several relatively low resolution N-body 
simulations with different random initial density fields.  These simulations are 
stopped at different redshifts and a group finding algorithm is used to find 
the locations of halos.  Halos of the form of equation~(\ref{halo}) are then put 
in, random lines of sight are propagated through the simulation and the magnification 
of each SN is calculated using equation~(\ref{kappa}).

Figure~\ref{fig:1} shows the results of some of these simulations.
It is clear that the medians of the distributions are all less than $\kappa=0$.  Most 
galaxies are demagnified.  It is also evident that $\kappa\ll 1$ for the vast majority of SNe at this redshift.  This justifies assumption that the lensing is weak.

The minimum possible value of $\kappa$ occurs when the line of 
sight does not intersect any halo.  This might be called the ``empty beam'' 
approximation for the angular size distance although in this case not all of the 
mass is in the halos so the beam is not truly empty.  This minimum $\kappa$ is 
represented by the vertical lines in figure~\ref{fig:1}.  It can be seen that in 
$\Lambda$ models the minimum $\kappa$ is smaller and the distribution extends to 
lower $\kappa$ than in open models.  In addition it is more likely in a $\Lambda$ 
model that there will be a close-encounter with a halo.  A result of these 
two things is that the variance of $P(\kappa)$ is larger in a $\Lambda$ model 
than in an open model, all other things being equal.  The left panel in 
figure~\ref{fig:1} shows that the clustering of the halos does not make a great 
deal of difference in their lensing properties in these models.  This means that 
the important scales for the lensing of point sources are $\simlt 1\mpc$, a range 
not accessible in shear maps except in the atypical cases of clusters.

There is information in probability distribution function of $\kappa$ (the 
$\kappa$-p.d.f) not only on the background cosmology (which can be gained in other, 
better ways) and the 
density profile of the halos, but also on the nature of dark matter.  I have 
implicitly assumed in this paper that the dark matter can be treated as a smooth 
density distribution for the purposes of lensing.  If the dark matter is in planet 
or stellar mass objects the $\kappa$-p.d.f. will change drastically.  The peak 
of the distribution will be taller and narrower and very close to the minimum 
$\kappa$.  For more on this use of the $\kappa$-p.d.f. see Metcalf (1998).

\section{Simulations of Observations}
The goal here is to predict how well future observations of SNe could 
constrain parameters of the halo model.  Or conversely, how much and 
what kind of data will be needed to make a good measurement of the halo parameters.

The method I propose here is to cross-correlate properties of observed foreground 
galaxies with the corrected peak luminosity of the SNe.  Using photometric redshifts 
and a trial Tully-Fisher relation a trial $\kappa$ for each SNe can be calculated.  
After a number of SNe have been observed the halo model can be adjusted to best 
fit the data.  The model outlined in the previous section has the following set of 
free parameters: $\{V_f,r_s,\gamma,\beta,\Omega_h\}$, if $L_f$ is taken to be fixed.  
Some of these would normally be held fixed as others are adjusted to fit the data.

The major sources of uncertainty are in the photometric redshifts, the scatter in the 
Tully-Fisher relation and the scatter in the intrinsic peak luminosities of the 
SNe.  Of these the latter by far dominates the uncertainty.

Monte Carlo calculation is done to assess how well observations are likely to 
constrain parameters.  A set of $N$ SNe is created with their associated lensing 
halos.  Many realizations of random errors are then generated to make simulated 
data sets.  For each data set a set of best fitting parameters is found.  The 
spread of these recovered parameters about the set that was used as input is a measure 
of an experiment's uncertainties so far as the input model is realistic.

\begin{figure}
\vspace*{0cm}
\centering\mbox{\psfig{figure=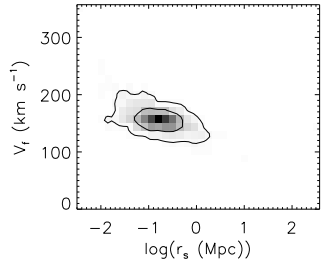,height=4.0cm}}
\vspace*{-0.5cm}
\caption[]{A two dimensional histogram representing the results of a Monte Carlo 
simulation for $\Omega_o=0.3$, $\Omega_h=0.18$, $\Omega_\Lambda=0.7$ with $\gamma=1$ 
halos.  The shading represents how many realizations of the observation fell within 
that region.  The contours contain $\% 63$ and $\% 95$ of the realizations.}
\label{fig:2}
\end{figure}

\begin{figure}
\vspace*{0cm}
\centering\mbox{\psfig{figure=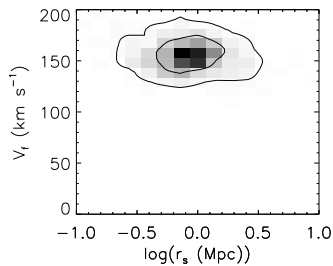,height=4.0cm}}
\vspace*{-0.5cm}
\caption[]{The same as figure~\ref{fig:2} only with $\Omega_h=0.29$ and $\gamma=3$.}
\label{fig:3}
\end{figure}

Figures~\ref{fig:2} and \ref{fig:3} shows the results of two such simulations.  
Here the constraints on the combination of $V_f$ and $r_s$, the scale size of the 
halos, are shown if 300 SNe at $z=1$ are observed.  The variance in the SNe's peak 
magnitude is 
taken to be $\sigma_m=0.12$.  There is a bias in the recovered parameters, 
but it is small in comparison to the uncertainties.
The main difference between these two cases is in $\Omega_h$ - $\gamma$ is not 
strongly constrained.

Simulations with 100 SNe at $z=0.9$ and $\sigma_m=0.16$ still show a 
significant detection of lensing and a constraint on $\log(r_s)$ of $\sigma=0.44$.  
Allowing the scale length to be correlated with the 
galaxies' luminosity, like for example $r_s \propto \sqrt{L}$, increases the 
dispersion in $\kappa$, and thus the signal, because more mass is put into rarer 
bright objects.


\section{Discussion}
\label{discussion}
How many SNe are there?  If the comoving type Ia SN rate is constant it is 
expected that there are $500-2000\mbox{ SNe}/\deg^2/\yr$ with $z<1.5$.  The 
larger end of this range is for the low $\Omega_o$ models.  The SN needs to be 
caught within about a one week (times $1+z$) window to be useful.  This 
gives $\sim 15-80\mbox{ SNe}/\deg^2$ below $z=1.5$ at any given time.  In addition 
since the star formation rate goes up with redshift it stands to reason that the 
SN rate goes up as well. This could increase the SNe rate by as much as a factor of 
5 (see Ruiz-Lapuente P. \& Canal R., (1998), Sadat {\it et al.} (1998) and others 
for discussions of the Ia SN rate).  The SNe are certainly out there.

The detection of type Ia SNe at $z=0.97$ has already been demonstrated although this 
SN was not spectroscopically identified as a type Ia (Garnavich, {\it et al.} 1998).  
The spectroscopic identification of the SNe as type Ia is the limiting factor in 
pushing to higher redshift right now.
In addition to high redshift SNe, low redshift SNe will be needed to accurately 
determine the distribution of SN peak luminosities and hopefully to improve the 
techniques used to correct the peak luminosity.

With the simple model presented here it has been demonstrated that with 100 or more 
SNe at $z \simgt 0.9$ something meaningful can be said about the nature of dark 
matter halos.  Higher redshift would help of course, but even with $z=0.9$, 
$\sigma_m=0.16$ and $N_{SN} \simlt 100$ lensing can be detected if most of the 
mass in the universe is within a few Mpc of a galaxy and is made of small particles.

There are a number of large field of view CCD cameras coming on line in the next 
few years.  Combining searches for lensing through shear with searches for SNe 
using these cameras would be mutually beneficial since the lensing of SNe provides 
information on scales not accessible in shear maps.

It should also be mentioned that it may also be possible to detect 
the lensing of SNe without resorting to correlations between matter 
and light.  This subject is addressed in Metcalf (1998).

\section*{Acknowledgments}

Thanks go to J. Silk and A. Riess for very helpful conversations, to J. Baker and M. Craig for 
providing the N-body code and to 
NASA for financial support.

\end{document}